\documentclass[prb,aps,showpacs,prb,amsmath,floatfix,twocolumn]{revtex4}
\usepackage{graphicx}
\usepackage{dcolumn}
\usepackage{bm}

\begin{document}
\title{Prediction of femtosecond oscillations in the transient current of
a quantum dot in the Kondo regime}
\author{A. Goker$^{1,2}$, Z. Zhu$^{2}$, U. Schwingenschl\"ogl$^{*}$ $^{2}$, and A. Manchon$^{2}$}

\affiliation{$^{1}$ Department of Physics, Bilecik University, 11210, Gulumbe, Bilecik, Turkey}

\affiliation{$^{2}$ KAUST, PSE Division, 23955-6900 Thuwal, Kingdom of Saudi Arabia}

\date{\today}


\begin{abstract}
We invoke the time-dependent non-crossing approximation 
in order to study the effects of the density of states 
of gold contacts on the instantaneous conductance of a 
single electron transistor which is abruply moved into 
the Kondo regime by means of a gate voltage. For an 
asymmetrically coupled system, we observe that the 
instantaneous conductance in the  Kondo timescale exhibits 
beating with distinct frequencies, which are proportional
to the separation between the Fermi level and the sharp 
features in the density of states of gold. Increasing the 
ambient temperature or bias quenches the amplitude of the 
oscillations. We attribute the oscillations to interference 
between the emerging Kondo resonance and van-Hove singularities 
in the density of state. In addition, we propose an experimental 
realization of this model.    
\end{abstract}



\thispagestyle{headings}

\maketitle

There is a growing need to provide a quantitative 
description of the sudden switching behaviour of 
single electron transistors, since the continuous
shrinking of conventional MOSFET transistors 
\cite{semiconductor} and developments in 
state-of-the-art nanotechnology experiments 
suggest they may one day constitute the building 
blocks of organic computers. Real-time electron 
dynamics in these devices have also profound 
implications for quantum computing \cite{ElzermanetAl04Nature} 
and the realization of an electron analog of a 
single-photon gun \cite{FeveetAl07Science}.

The effect of sudden perturbation in the form of step-like 
switching of the gate or bias voltage has been studied 
thoroughly \cite{NordlanderetAl99PRL,PlihaletAl00PRB,
SchillerHershfield00PRB,MerinoMarston04PRB} and it has 
been unambiguously shown that the resulting transient 
current exhibits different time scales \cite{PlihaletAl05PRB,
AndersSchiller05PRL,AndersetAl06PRB,IzmaylovetAl06JPCM}. 
The initial fast non-Kondo timescale is characterized by the 
reshaping of the broad Breit-Wigner resonance located around 
the impurity level, while the sharp Kondo resonance pinned 
to the Fermi level reaches a metastable state at the end of 
the much longer Kondo timescale. Subsequent investigations 
predicted that the asymmetric coupling of the dot to the 
contacts may induce interference between the Kondo resonance 
and the discontinuities in the density of states of the leads 
\cite{GokeretAl07JPCM}. Extension of the diagrammatic Monte Carlo 
method \cite{GulletAl08EPL} to impurities out of equilibrium 
\cite{WerneretAl09PRB} verified the transient current's 
dependency on the bandwidth of the contacts \cite{SchmidtetAl08PRB}.

Ramifications of the band structure of contacts in time-dependent 
transport have been elucidated previously for prototypical systems 
by directly solving the Green's functions in the time domain 
\cite{ZhuetAl05PRB,MaciejkoetAl06PRB,GokeretAl07JPCM}. Ab-initio 
calculations also suggest that the crystallographic orientation
of electrodes may influence transport properties \cite{Wangetal10JAP}.

The purpose of this letter is to provide a more realistic picture 
relevant to an actual experiment by using the real density of states 
of gold obtained from first principles calculations as input in a 
study of the transient current through a quantum dot asymmetrically 
coupled to gold leads. The dot level is abruptly switched such that 
the Kondo resonance is present in the final state. We analyze the 
effects of the density of states of the conduction electrons in the 
contacts, the position of the dot energy level, the asymmetries in 
the couplings, and the temperature on the instantaneous current.

The physics of this system is captured sufficiently by a single 
impurity Anderson Hamiltonian which describes a single doubly 
degenerate level of energy $\epsilon_{dot}$ attached to continuum 
electron baths. There is a one-to-one mapping between this model 
and the Kondo model via Schriefer-Wolff transformation for the 
parameter range we will be interested in this paper. The auxiliary 
boson transformation is performed for the Anderson Hamiltonian. 
In this method, the electron operator on the impurity is replaced 
by a massless boson operator and a pseudofermion operator. The 
$U\rightarrow\infty$ limit is achieved by restricting the sum of 
the number of bosons and pseudofermions to unity. The Hamiltonian 
is then converted into
\begin{eqnarray}
H(t)&=&
\sum_{k\alpha\sigma}\left [\epsilon_{k}n_{k\alpha\sigma}+
V_{\alpha}(\varepsilon_{k\alpha},t)c_{k\alpha\sigma}^{\dag}
b^{\dag}f_{\sigma}+{\rm H.c.} \right]+ \nonumber \\
& & \sum_{\sigma}\epsilon_{dot}(t)n_{\sigma},
\end{eqnarray}
where $f_{\sigma}^{\dag}$ $(f_{\sigma})$ and $c_{k\alpha\sigma}^{\dag}$ 
$(c_{k\alpha\sigma})$ with $\alpha=L,R$ create (annihilate) an electron 
of spin $\sigma$ in the dot and in the left (L) and right (R) gold leads, 
respectively. Moreover, $n_{\sigma}$ and $n_{k\alpha\sigma}$ are the 
corresponding number operators, $V_{\alpha}$ are the tunneling amplitudes 
for the left and the right leads, and $b^{\dag}$ $(b)$ creates (annihilates) 
a massless boson in the impurity.  

Assuming that the hopping matrix elements have no explicit time 
dependence, the coupling of the quantum dot to the contacts can be 
parametrized as $\Gamma_{L(R)}(\epsilon)=\bar{\Gamma}_{L(R)} \xi_{L(R)}(\epsilon)$,
where $\bar{\Gamma}_{L(R)}=2\pi|V_{L(R)}(\epsilon_f)|^2$ is a constant
and $ \xi_{L(R)}(\epsilon)$ is the density of states function. In 
order to describe the density of states of the gold contacts 
accurately, we employ ab-initio calculations.

\begin{figure}[t]
\centerline{\includegraphics[angle=0,width=8.5cm]{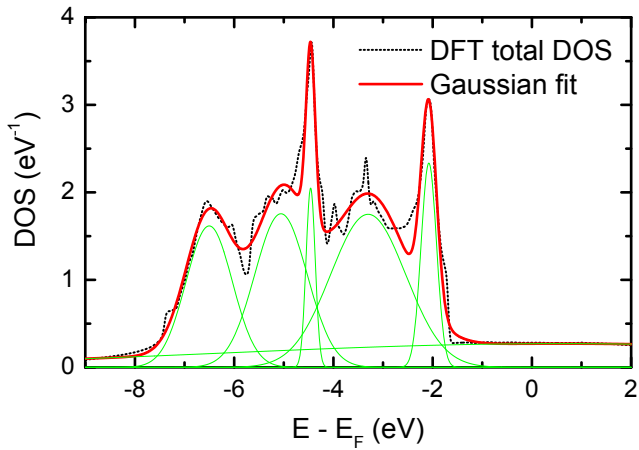}}
\caption{
The total density of states of gold is shown by a black dashed 
curve as a function of the separation from the Fermi level. The 
red curve corresponds to the best fit to the ab-inito data by a 
linear combination of Gaussian functions, which are represneted 
by green curves.}
\label{Fig1}
\end{figure}

The ab-initio calculation of the gold density of states is based 
on density functional theory. We make use of the well-established 
WIEN2K package \cite{Blahaetal01Book,us}, in which the full-potential 
linearized augmented plane wave method is implemented with a dual-basis 
set. Specifically, the gold core states are defined by the electronic
configuration Kr $4d^{10}4f^{14}5s^{2}$, whereas the valence states 
comprise $5p$, $5d$, $6s$, and $6p$ orbitals. The face-centered cubic 
unit cell of gold (space group $Fm\bar{3}m$, lattice constant $a$ = 4.080 \AA) 
with 4 atoms per cell is used in the calculation. For the exchange-correlation 
potential the generalized gradient approximation within the Perdew, Burke, 
and Ernzerhof (GGA-PBE) parametrization is adopted \cite{Perdewetal96PRL}.
The radius of the gold muffin-tin sphere is set to 2.5 Bohr radii. 
Moreover, the plane wave cut-off for the scalar relativistic basis 
function is given by $R_{mt}K_{max}=8$ and $l_{max}=10$. Integrations 
in the reciprocal space for self-consistent field cycles apply the 
tetrahedron method and 752 k-points in the irreducible wedge of the 
Brillouin zone. The resulting density of states (DOS) is shown in 
Fig.~\ref{Fig1}.

In order to represent this complicated curve in a functional 
form, we use a fitting procedure with a linear combination 
of Gaussian functions given by 
\begin{equation}
\rho(\epsilon)=\sum_i \rho_i(\epsilon),
\end{equation}
where
\begin{equation}
\rho_i(\epsilon)=\frac{\alpha_i}{\zeta_i \sqrt{0.5\pi}}\exp(-2(\frac{\epsilon-\epsilon_i}{\zeta_i})^2).
\end{equation}
The outcome of the fitting procedure, that involves
six different Gaussians of varying linewidth and peak 
position, is also shown in Fig.~\ref{Fig1}. The number 
of Gaussians corresponds to the number of gold states 
in the vicinity of the Fermi level (five $d$ orbitals 
and one $s$ orbital). The resulting best fit parameters
are summarized in Table~\ref{parameters}. Throughout the 
rest of this letter, we will switch to atomic units, where 
$\hbar=k_B=e=1$, and perform our calculations accordingly.

\begin{table}[ht]
\caption{Fitting parameters for 6 different Gaussians
used to replicate the density of states of gold (in eV).}
\centering
\begin{tabular}{c | c | c | c | c | c | c}
 & 1 & 2 & 3 & 4 & 5 & 6 \\
\hline
$\epsilon_i$ & -6.503 & -5.057 & -4.459 & -3.300 & -2.077 & 0.557 \\
\hline
$\zeta_i$ & 0.934 & 1.055 & 0.185 & 1.513 & 0.310 & 13.474 \\
\hline
$\alpha_i$ & 1.892 & 2.321 & 0.475 & 3.317 & 0.907 & 4.580 \\
\hline
\end{tabular}
\label{parameters}
\end{table}

We invoke the well tested non-crossing approximation (NCA) 
to obtain the pseudofermion and slave boson self-energies. 
NCA gives reliable results for dynamical quantities except 
for very low temperatures or finite magnetic field. We stay 
away from both these regimes. We solve the resulting real-time 
coupled integro-differential Dyson equations for the retarded 
and lesser Green's functions in a discrete two-dimensional grid. 
A technical description of our implementation has been published 
elsewhere \cite{ShaoetAl194PRB,IzmaylovetAl06JPCM}.

The net current in the electrical circuit can be derived 
from the Green's functions $G_{pseu}^{<(R)}(t,t')$ and
$B^{<(R)}(t,t')$. We will denote the net current by
$I(t)=I_L(t)-I_R(t)$, where $I_L(t)$ $(I_R(t))$ represents 
the net current from the left (right) contact through the
left (right) barrier to the dot. The general expression
for the net current \cite{JauhoetAl94PRB} can be rewritten 
using pseudofermion and slave boson Green's functions 
\cite{GokeretAl07JPCM}, finally leading to
\begin{eqnarray}
& & I(t) =-2(\bar{\Gamma}_{L}-\bar{\Gamma}_{R})\textit{Re} \left (\int_{-\infty}^{t} dt_1
\xi_{o}(t,t_1)h(t-t_1)\right)+\nonumber \\
& & 2\bar{\Gamma}_{L} Re \left (\int_{-\infty}^{t} dt_1 
(\xi_{o}(t,t_1)+\xi_{u}(t,t_1)) f_{L}(t-t_1) \right)- \nonumber \\
& & 2\bar{\Gamma}_{R} Re \left (\int_{-\infty}^{t} dt_1
(\xi_{o}(t,t_1)+\xi_{u}(t,t_1)) f_{R}(t-t_1)\right),\nonumber \\
\label{final}
\end{eqnarray}
with $\xi_{o}(t,t_1)=G_{pseu}^{<}(t,t_1)B^{R}(t_1,t)$ as 
well as $\xi_{u}(t,t_1)=G_{pseu}^{R}(t,t_1)B^{<}(t_1,t)$. 
In Eq.~(\ref{final}), $f_L(t-t_1)$ and $f_R(t-t_1)$ are 
the convolutions of the density of states function with the 
Fermi-Dirac distributions of the left and right contacts, 
respectively, while $h(t-t_1)$ is the Fourier transform of 
the density of states \cite{GokeretAl07JPCM}. The conductance 
$G$ is given by the current divided by the bias voltage 
$V$. The subsequent time dependent conductance results are
computed by Eq.~(\ref{final}). We will be referring to 
$\eta=\frac{\bar{\Gamma}_{L}}{\bar{\Gamma}_{tot}}$, where 
$\bar{\Gamma}_{tot}=\bar{\Gamma}_{L}+\bar{\Gamma}_{R}$, 
as the asymmetry factor.

The Kondo effect is a many-body resonance arising at low 
temperatures due to a spin singlet formed from the 
hybridization of the net free spin inside the dot with the 
continuum electrons in the contacts. It originates from 
the seminal work of Jun Kondo \cite{Kondo64PTP}, where he 
discovered that a divergence in the perturbation series
of tunnel couplings yields resistance enhancements in 
metals containing magnetic impurities. Its hallmark is a 
sharp resonance pinned to the Fermi levels of the contacts 
in the dot density of states. The broadening of the Kondo 
resonance is given by an energy scale $T_K$ (Kondo temperature), 
defined as
\begin{equation}
T_K \approx \left(\frac{D\Gamma_{tot}}{4}\right)^\frac{1}{2}
\exp\left(-\frac{\pi|\epsilon_{\rm dot}|}{\Gamma_{tot}}\right).
\label{tkondo}
\end{equation}
Here, $D$ is an energy cutoff, equal to half the bandwidth
of the conduction electrons, and $\Gamma_{tot}$ is the value 
of the total coupling between the dot and the contacts
$\Gamma_{tot}(\epsilon)$ at $\epsilon=\epsilon_F$.
  
\begin{figure}[t]
\centerline{\includegraphics[angle=0,width=8.5cm]{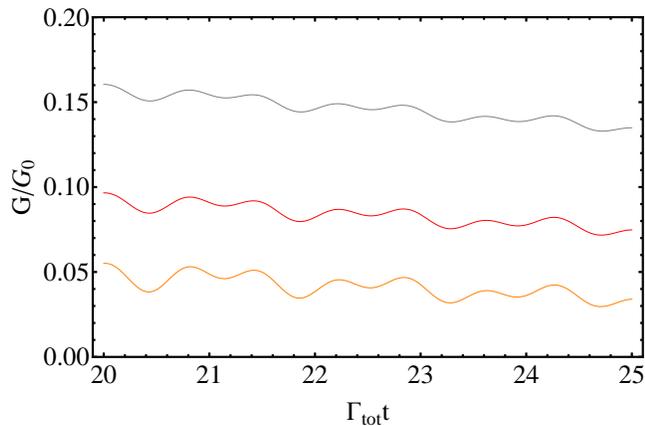}}
\caption{
The grey, red and orange curves (from top to bottom) show the 
instantaneous conductance versus time in the Kondo timescale 
after the dot level has been switched to its final position for 
asymmetry factors of 0.85, 0.9 and 0.95, respectively, at 
$T=0.009\Gamma_{tot}$ and $V=T_K$ for constant $\Gamma_{tot}$.}
\label{Fig2}
\end{figure}

In the following, we investigate the instantaneous 
conductance for a system in which the dot level is 
abruptly shifted from $\epsilon_1=-5\Gamma_{tot}$ to 
$\epsilon_2=-2\Gamma_{tot}$ where $\Gamma_{tot}=$0.8 eV
at $t=0$ by a gate voltage, resulting in a transition 
from a non-Kondo state ($T_K \ll T$) to a Kondo state. 
In the final state, we infer $T_K=0.0025\Gamma_{tot}$ 
from Eq.~(\ref{tkondo}). In the initial short timescale 
associated with charge fluctuations, the conductance 
reaches a maximum for large asymmetry factors and then 
starts to decay, confirming previous studies \cite{GokeretAl07JPCM}. 
Fig.~\ref{Fig2} shows the behaviour of the instantaneous 
conductance in the long Kondo timescale after the dot 
level has been switched to its final position. The 
instantaneous conductance exhibits a complex ringing 
behaviour and the amplitude of the oscillations diminishes 
with decreasing asymmetry factor, completely disappearing 
for symmetric coupling. This is simply because the interference 
between the left contact and the Kondo resonance is out of 
phase with the interference between the right contact and 
the Kondo resonance due to opposite signs in Eq.~(\ref{final}). 
The amplitudes of these two interference processes are 
equal for symmetric coupling and the oscillations cancel out.

\begin{figure}[t]
\centerline{\includegraphics[angle=0,width=8.5cm]{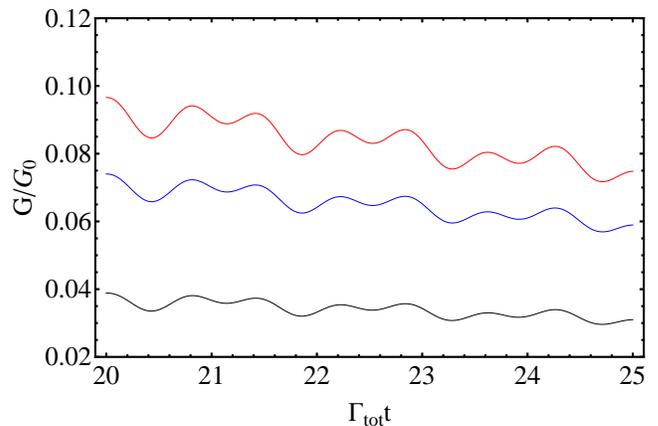}}
\caption{
The red, blue and black curves (from top to bottom) represent the 
instantaneous conductance versus time in the Kondo timescale after 
the dot level has been switched to its final position for an
asymmetry factor of 0.9 at $T=0.009\Gamma_{tot}$ when the 
bias is equal to $V=T_K$, $V=4T_K$ and $V=8T_K$, respectively, 
for constant $\Gamma_{tot}$.}
\label{Fig3}
\end{figure}

The frequencies taking part in the oscillations can 
be extracted by a Fourier transform of the time 
dependent conductance. It turns out that there exist 
two distinct frequencies, $\omega_1$ and $\omega_2$, 
which give rise to a beating behaviour with envelope 
and carrier frequencies of $\omega_1-\omega_2$ and 
$\omega_1+\omega_2$, respectively. We find 
$\omega_1=2.15\omega_2$, which corresponds to the
ratio of the distances of the peaks at $-2.08$ eV and 
$-4.46$ eV in Fig.~\ref{Fig1} from the Fermi level. 
This result demonstrates that van-Hove singularities 
can be probed by the transient current in the Kondo regime.

The effect of finite bias on the instantaneous 
conductance is depicted in Fig.~\ref{Fig3}. The 
influence of finite bias is twofold. First, it 
reduces the amplitude of the oscillations. Second, 
the decay rate of the oscillations, defined as the 
inverse of the time it takes the oscillations to die out,
increases, indicating that the oscillations are inherently
related to the formation of the Kondo resonance. It 
should be noted that the split Kondo peak oscillations
arising at finite bias have no discernible effect on
these results since their frequency is two orders of
magnitude smaller.

\begin{figure}[b]
\centerline{\includegraphics[angle=0,width=8.5cm]{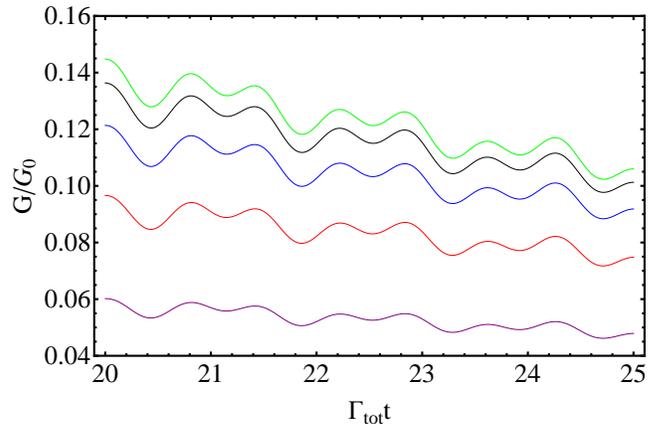}}
\caption{
The green, black, blue, red and purple curves (from top to bottom) represent the 
instantaneous conductance versus time in the Kondo timescale after the dot 
level has been switched to its final position for an asymmetry factor of 0.9
at $T=0.0015\Gamma_{tot}$, $T=0.0030\Gamma_{tot}$, $T=0.0060\Gamma_{tot}$, 
$T=$ $0.0090\Gamma_{tot}$ and $T=0.0150\Gamma_{tot}$, respectively, for constant 
$\Gamma_{tot}$ with a bias of $V=T_K$.}
\label{Fig4}
\end{figure}

Finally, the effect of ambient temperature is displayed
in Fig.~\ref{Fig4}. Lowering the ambient temperature 
influences the time dependent conductance results in 
the following ways. First, the decay rate of the oscillations 
decreases. Second, the amplitude of the oscillations 
starts increasing, but saturates when the temperature 
approaches the Kondo temperature $T_K$. Lowering the 
temperature below $T_K$ does not alter the amplitude anymore. 

Based on these observations and the fact that discontinuities
in the density of states of the contacts can induce
interference with the Kondo resonance for an asymmetrically
coupled system \cite{GokeretAl07JPCM}, we suggest that the 
beating behaviour in the time dependent conductance is a 
result of interference between the emerging Kondo resonance 
at the Fermi level and the sharp features in the density of 
states of gold located at $-2.08$ eV and $-4.46$ eV. 
Since the electrons in the contacts are assumed to 
be non-interacting, the density of states of the contacts 
is time independent. It always looks like Fig.~\ref{Fig1}.
As a result, sharp features in it are also static.
On the other hand, the Kondo resonance is part of the 
density of states of the dot. Consequently, its formation 
is time-dependent and dynamical. This naturally leads to 
persistence of the beating behaviour until the Kondo resonance 
is fully formed. The ratio of $\omega_1$ to $\omega_2$ 
supports this scenario. Moreover, the reduction of the 
oscillation amplitudes with increasing source-drain bias 
is due to the fact that the Kondo resonance starts getting 
destroyed and the interference vanishes. The saturation 
of the oscillation amplitudes below $T_K$ is a further testimony 
to the proposed mechanism, as the Kondo resonance is best 
developed below this scale. Hence, the interference strength 
stays the same. We want to note that the validity of our
results in transient regime can be checked by considering
a second system which has a smaller $T_K$. This can be
accomplished by taking a slightly lower final dot level
$\epsilon_{dot}$. The time dependent conductance curves
overlap for both systems as a function of $T_K t$. This
proves that our calculations obey universality which is
a hallmark of Anderson model.

In conclusion, we have studied the transient current in a 
single electron transistor consisting of gold contacts in 
response to an abrupt switching of its dot level. We have 
used the density of states of gold, as obtained from ab-initio 
electronic structure calculations, as input in a many-body 
calculation. We find that an asymmetrically coupled system 
exhibits complex oscillations in the Kondo timescale. The two 
distinct frequencies that give rise to the observed beating 
behaviour are found to be proportional to the separation between 
the Fermi level and the two sharp features in the density of 
states. We interpret this as an interference between the emerging 
Kondo resonance and the van-Hove singularities. We note
that more frequencies may mix with the two dominant frequencies
in an actual experiment due to additional features in the gold
density of states, see Fig.~\ref{Fig1}. However, their effects on
the reported findings would not be discernible as the amplitudes
are negligible as compared to those captured by our fitting. 

Finally, we feel that the situation discussed in this letter can
be realized with today's experimental technology, since 
ultrafast pump-probe experiments can capture the picosecond 
timescale \cite{Teradaetal10JPCM}. This is close to the time 
needed for forming the Kondo resonance \cite{NordlanderetAl99PRL}. 
It would be greatly desirable to confirm our predictions experimentally.

\end{document}